\documentclass[aps,prb,twocolumn,groupedaddress,epsfig,psfig,showpacs]{revtex4-1}

\usepackage{graphicx}

\def\STO {SrTiO$_{3}$}
\def\LAO {LaAlO$_{3}$} 

\begin{document}

\title{Investigation of the In-Gap Electronic Structure of \LAO\ - \STO\ Heterointerfaces by Soft X-ray Spectroscopy}

\author{A. Koitzsch$^{1}$, J. Ocker$^{1}$, M. Knupfer$^{1}$, M. C. Dekker$^{1}$, K. D\"orr$^{1}$, B. B\"uchner$^{1}$, and P. Hoffmann$^{2}$}

\affiliation
{$^{1}$Institute for Solid State Research, IFW-Dresden, P.O.Box 270116, D-01171Dresden, Germany\\
$^{2}$Helmholtz-Zentrum Berlin, BESSY, Albert-Einstein-Str. 15, D-12489 Berlin, Germany}

\date{\today\\}

\begin{abstract}
We investigated \LAO\ - \STO\  heterointerfaces grown either in oxygen rich or poor atmosphere by soft x-ray spectroscopy. Resonant photoemission across the Ti L$_{2,3}$ absorption edge of the valence band and Ti 2p core level spectroscopy directly monitor the impact of oxygen treatment upon the electronic structure. Two types of Ti$^{3+}$ related charge carriers are identified. One is located at the Fermi energy and related to the filling of the \STO\  conduction band. It appears  for low oxygen pressure only. The other one is centered at E$_{B}$ $\approx$ 1 eV and independent of the oxygen pressure during growth. It is probably due to defects. The magnitude of both excitations is comparable. It is shown that low oxygen pressure is detrimental for the Ti - O bonding. Our results shed light on the nature of the charge carriers in the vicinity of the \LAO\  - \STO\  interface. 

\end{abstract}

%\pacs{71.18.+y, 74.25.Jb, 74.72.-h}

\maketitle

\section{Introduction\\}
Fascinating and counterintuitive phenomena have been observed at the interface of \STO\  and \LAO. The most important is the appearance of metallic conductivity between two firm insulators \cite{Ohtomo}. Intriguingly the interface remains insulating as long as no more than three layers of \LAO\  are deposited on top of \STO\  but switches to metallic with the fourth layer \cite{Thiel2006,Huijben2006}. Subsequently, also superconductivity was found below T$_{C}$ = 0.2 K \cite{Reyren2007} and even magnetism \cite{Brinkman2007}. The picture of the “polar catastrophe” has been invoked early on to explain the metallic conductivity: stacking of LaO$^{+}$ and AlO$_{2}$$^{-}$ layer by layer on top of \STO\ leads to a divergent electrostatic potential which is compensated by a charge transfer of 0.5 e$^{-}$ per unit cell to the interface for the case of LaO deposited on TiO$_{2}$. These charge carriers are hold responsible for metallicity at the interface. 

The polar catastrophe is conceptually simple and explains elegantly the change of groundstates for three vs four layers of \LAO. However, experimental evidence for the importance of oxygen vacancies as effective doping mechanism has been presented \cite{Kalabukhov2007, Siemons2007a}. In fact the observed sheet carrier density for heterostructures grown in oxygen poor atmosphere (p$_{O2}$ = 10$^{-6}$ mbar) assumes values several orders of magnitudes larger than what is expected at maximum for e/2 charge transfer for the polar catastrophe. On the other hand, for samples grown under higher oxygen partial pressure the charge density significantly decreases. These observations support the view that oxygen vacancies are essential for the properties of the interface. Depending on growth conditions they may completely dominate the low energy properties or - at least - compete with the polar catastrophe.
A third important concept for the description of the \LAO\ - \STO\ heterostructure is  intermixing of the chemical species, which tends to mitigate polar discontinuities or to invoke a (La, Sr)TiO$_{3}$ layer at the interface \cite{Willmott2007, Kalabukhov2009, Qiao2010, Siemons2010}. 

Photoemission spectroscopy is sensitive to all of these three mechanisms, polar catastrophe, oxygen vacancies and intermixing, and has been applied previously to the \LAO\ - \STO\ system.
Segal et al. did not observe the consequences of strong internal fields connected to the \LAO\ layer such as line broadening and shifting \cite{Segal2009}. 
Metallic charge carriers arising from electronic reconstruction or oxygen vacancies form a Fermi edge in photoemission experiments as soon as their concentration is detectable. 
For the \LAO\ - \STO\ system the intensity at E$_{F}$ has been studied as a function of p$_{O2}$ and found to decrease with p$_{O2}$  increase for standard ultraviolet photoemission \cite{Siemons2007b}.
Interestingly, no Fermi edge signal and no other in gap states were found for Molecular Beam Epitaxy (MBE) grown metallic samples even for the Ti L resonance (which is known to greatly enhance the Ti 3d derived density of states) from which upper limits for the charge concentration were deduced \cite{Yoshimatsu2008}. In contrast, a recent study on pulsed laser deposition (PLD) - grown samples did find in-gap states reaching up to E$_{F}$, whose intensity depends on the number of \LAO\ layers \cite{Drera2011}.
Cation intermixing has been studied by monitoring the intensity profiles of suitable core levels excitations as a function of emission angle \cite{Qiao2010, Chambers2010}. These profiles are inconsistent with an abrupt interface.

It is fair to say that no coherent picture for the spectroscopic investigations of the \LAO\ - \STO\ system has emerged yet. Here we study the in-gap states with particular attention to the influence of the oxygen pressure during growth. The latter is apparently a decisive parameter for the \LAO\ - \STO\ heterostructures and has a direct and well documented impact on the transport properties, e.g. it tunes the electrical conductivity over many orders of magnitude. 
We consider two limiting cases: samples grown under  low (4.5  $\cdot$  10$^{-6}$ mbar) and high (5  $\cdot$  10$^{-3}$ mbar) oxygen atmosphere. The goal of this study is to monitor the effect of the latter on the electronic structure. To this end we recorded detailed photoemission spectra of the valence band and Ti-states. 

The paper is structured as follows: In section II the experimental procedures are described as well as sample preparation and characterization. Section III presents the results from valence band PES and from the core levels. In section IV the results are discussed.

\section {Experimental}

Experiments have been carried out at the Helmholtz – Zentrum – Berlin (BESSY) at undulator beamline UE 52. Photoemission measurements have been performed using a Scienta R4000 photoelectron analyzer. The total energy resolution has been set  to 260 meV for 440 eV photon energy and 620 meV for 1200 eV. 
X-ray absorption has been measured by monitoring the drain current. 

\LAO\ - \STO\ heterostructures with an conducting LaO - TiO$_{2}$ interface were fabricated on atomically flat TiO$_{2}$ terminated \STO\ single crystals by pulsed laser deposition. In order to obtain TiO$_{2}$ terminated \STO\ substrates a cleaning and  BHF etching procedure was applied\cite{Koster1998}.  
The \STO\ single crystal, dried with argon gas was annealed at 950$\,^{\circ}\mathrm{C}$ in oxygen atmosphere for 3 hours, in order to recrystallize the substrate, hence forming atomically flat parallel terraces. During growth the thickness of each \LAO\ layer was controlled on an atomic scale by monitoring the intensity oscillations of the specular spot in reflection high-energy electron diffraction (RHEED). The samples were grown at 800$\,^{\circ}\mathrm{C}$ with a laser fluence of 2-3 J/cm$^{2}$, resulting in growth of approximately 0.7 \AA\  per pulse. No further oxidation step was performed after the deposition process. Sample OR (oxygen rich) was grown at 5.0 $\cdot$ 10$^{-3}$ mbar oxygen pressure and the thickness was fixed to 5 unit cells \LAO, whereas sample OP (oxygen poor) was grown at 4.5 $\cdot$ 10$^{-6}$ mbar with 6 unit cells. The samples have been cleaned by acetone rinsing before transfer to the UHV system. No further in situ cleaning procedure was applied before measurements.

The surface morphology of the heterostructures has been determined by atomic force microscopy. The step- and terrace structure of the substrate is preserved throughout the growth of the \LAO\ film for low and high oxygen partial pressures. For sample OP the \LAO\ surface is smooth and does not show any variations in microstructure.  Sample OR grown in higher oxygen pressure shows small islands of \LAO\ at the step edges. The islands are 1 unit cell high and appear independently of the growth temperature. 

\begin{figure}[h!]
\includegraphics[width=0.75\linewidth]{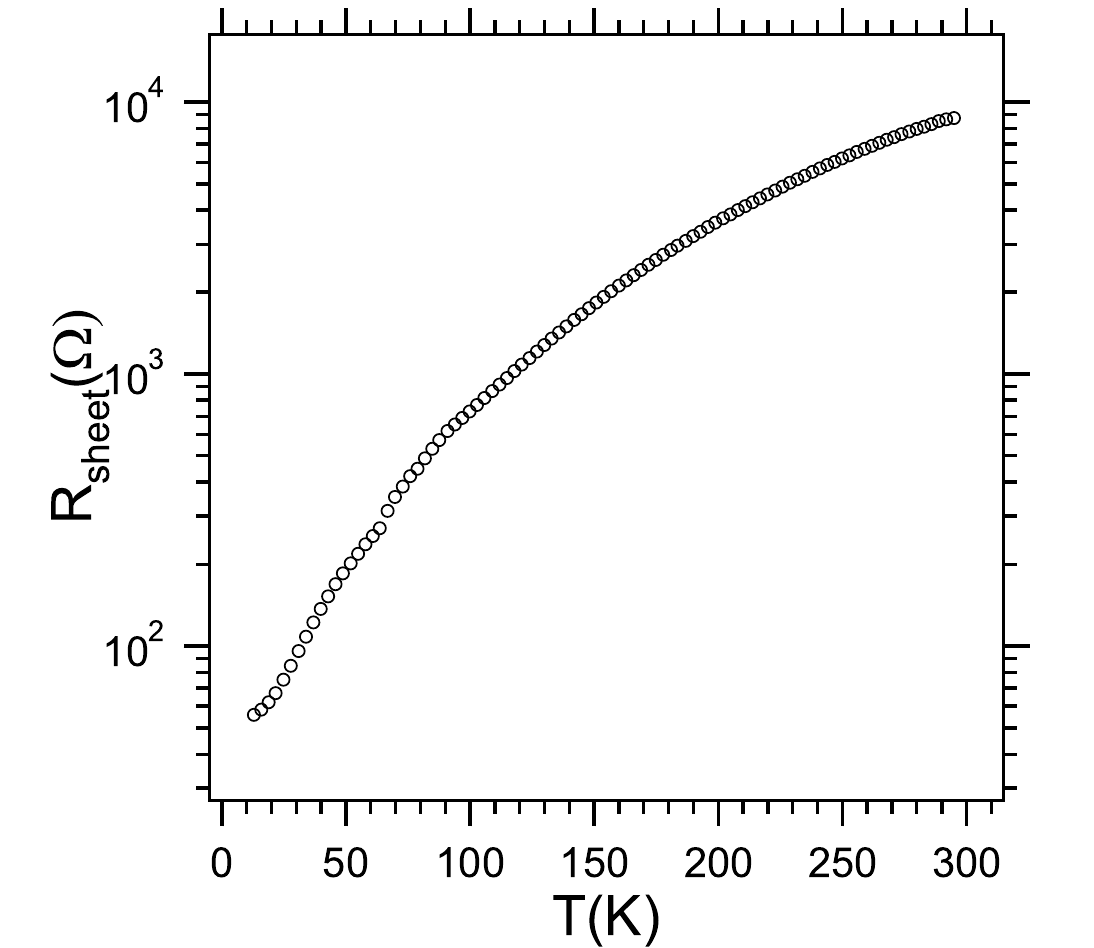}% Here is how to import EPS art
\caption{Sheet resistance of sample OP (grown in oxygen poor atmosphere).} \label{fig:Fig0}
\end{figure}

The conductivity of the samples has been checked in dependence on temperature using wire-bonded Al contacts. Fig. 1 shows the sheet resistance of the oxygen-poor (OP) sample which is metallic with a residual resistance ratio of R$_{sheet}$(300 K)/R$_{sheet}$(10 K) = 174. The magnitudes of R$_{sheet}$ and the resistance ratio fit well into the resistance characteristics recorded for various deposition pressures in Ref.~5, considering the pressure of 4.5 $\cdot$ 10$^{-6}$ mbar for the OP sample. The oxygen-rich (OR) sample proved to have a high sheet resistance with R$_{sheet}$ $>$ 10$^6$ $\Omega$ in the entire temperature range of 10 - 300 K, exceeding the range of measurable resistances of our setup. 

\section{Results}

\subsection{Valence Band}

To unravel the origin of metallic conductivity of the \LAO\ - \STO\ interface it is essential to investigate the electronic states at and near the Fermi energy. We start the presentation of the experimental results with valence band data of sample OP in Fig. 2.  Fig. 2a shows the valence band as a function of photon energy across the Ti L$_{2,3}$ absorption edge over a wide binding energy range capturing the shallow core levels which have been used for normalization. The Fermi energy (E$_{F}$) has been obtained by referencing to the Fermi edge of gold in electrical contact to the sample. However, small changes (in the order of 0.1 eV) occurred for the different photon energies across the edge energies, possibly related to residual charging. Therefore, we will not discuss absolute binding energies with this precision. 
In general the observed valence bands are a mixture of contributions from \STO\ and \LAO. However, near E$_{F}$ only \STO\ states are present. The band gap of pure \STO\ is 3.2 eV at room temperature and the conduction band minimum is located at 0.2-0.3 eV above E$_{F}$ \cite{Henrich1978}. With n-type doping the conduction band is occupied and the onset of the valence band shifts to higher binding energy. \LAO, on the other hand, has a larger band gap (5.6 eV). Conduction band minimum and valence band maximum of \LAO\ are each approximately 3 eV away from E$_{F}$, thus the close vicinity of E$_{F}$ is solely determined by \STO\ states.  

The bottom spectrum in Fig. 2a has been taken well before the onset of the Ti L$_{2,3}$ absorption edge, i.e. off- resonant.  For this off-resonant spectrum no Fermi edge is visible and no signature of in-gap states frequently observed for \STO\ and other Ti$^{4+}$ systems \cite{Henrich1978, Courths1989, Zhang1991, Prince1997,  Aiura2002, Takizawa2006, Ishida2008}. The main valence  band (E$_{B}$ = 3 - 9 eV)  is composed of a high and low binding energy region (V$_{1}$ and V$_{2}$). At higher binding energies shoulder structures (C and L) appear. Feature C at 11 eV is probably related to surface adsorbates, which are notorious in this energy region. The energy position of feature L coincides with a structure in the \LAO\ valence band.

The photon energy has been swept through the Ti L$_{2,3}$ absorption edge shown in Fig. 2c. The shape of the absorption spectra compares favorably with the result of a multiplet calculation for Ti$^{4+}$. The four prominent peaks of this spectrum correspond to the spin-orbit splitting of the Ti 2p level and the crystal field splitting in t$_{2g}$ and e$_{g}$ states for each of the spin-orbit components. In comparison, the Ti$^{3+}$ spectrum looks very different and can be easily distinguished (Fig. 2c, inset).  It is well known that the resonance photoemission process enhances the Ti 3d related density of states in the valence band. For a Ti$^{4+}$ system the Ti 3d bands are nominally empty, thus any resonance enhancement is a signature of Ti 3d - O 2p hybridization. 

\begin{figure}[h!]
\includegraphics[width=1\linewidth]{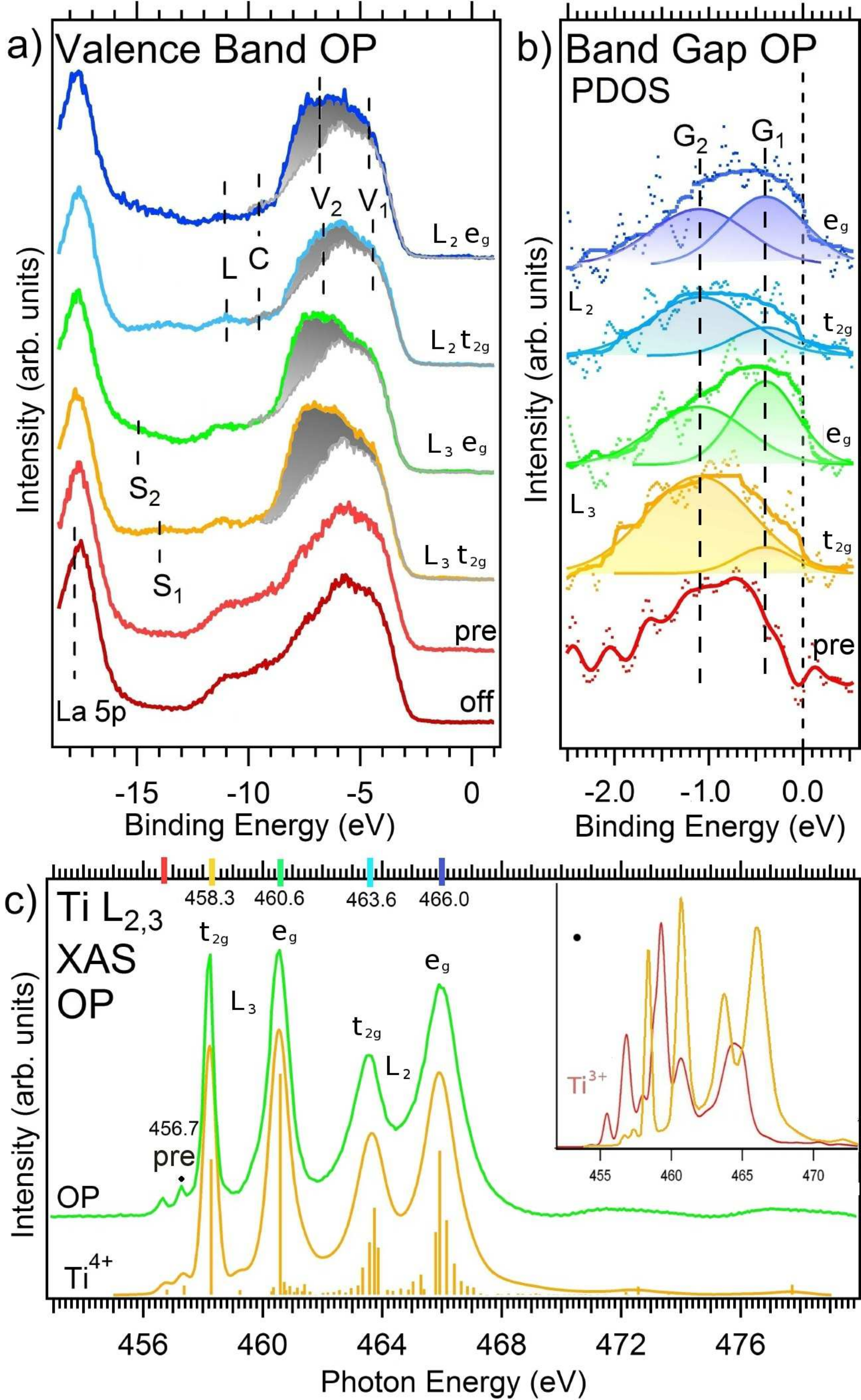}% Here is how to import EPS art
\caption{(Color online) Valence band photoemission of \LAO\ - \STO\ (sample OP, grown in oxygen poor atmosphere). a) Ti 2p $\rightarrow$ 3d resonant photoemission valence band obtained by varying the photon energy through the Ti L$_{2,3}$ absorption threshold. The grey area shows the enhancement of Ti 3d derived states by comparison with the off-resonant spectrum at h$\nu$  = 440 eV.  b) Difference spectra near E$_{F}$ obtained by subtracting the off resonant spectrum (h$\nu$ = 440 eV) from the on - resonant spectra. Thick lines are obtained from the original data by moderate smoothing. c) Ti L$_{2,3}$ absorption edge taken in total electron yield mode compared to a multiplet calculation \cite{Stavitski2010}. Color bars at the top indicate the photon energy of the valence band spectra. Inset shows a comparison of calculated L$_{2,3}$ Ti$^{4+}$ and Ti$^{3+}$  absorption edges.} \label{fig:Fig1}
\end{figure}

As the photon energy is tuned through the Ti L edge intensity enhancements occur in the valence band. We distinguish three characteristic regions: the low energy region near  E$_{F}$ (see expanded view in Fig. 2b), the enhancement of the main valence band (V$_{1}$ and V$_{2}$) and weak but discernible increase at higher binding energy around  E$_{B}$= 12 - 15 eV (satellites S$_{1}$ and S$_{2}$). The strongest enhancement occurs for the high binding energy region of the valence band (feature V$_{2}$), which is much more enhanced than the low binding energy region (V$_{1}$). This is in agreement with previous studies and has been interpreted as a consequence of predominantly bonding O 2p orbitals at the high binding part of the valence band with relatively strong Ti 3d admixture, whereas the region V$_{2}$ is dominated by nonbonding O 2p orbitals with little Ti 3d overlap \cite{Higuchi1998, Zhang1991, Ishida2008}. 

More interesting than the behavior of the main valence band is, however, the low energy region close to the Fermi energy. We observe clear intensity enhancement upon entering the resonance regime in the region of E$_{B}$ = 0 - 2 eV. In Fig. 2b we show the difference spectra obtained by subtracting the off-resonant spectra. The line shape of the remaining difference spectra can be described by two components, one centered close to the Fermi energy (G$_{1}$), the other one centered around E$_{B}$ = 1 eV (G$_{2}$). In particular feature G$_{1}$ is cut off by E$_{F}$ signaling the metallic nature of these charge carriers. These observations are similar to previous results for \STO\ \cite{Henrich1978, Ishida2008} and \STO\ heterostructures \cite{Takizawa2006, Siemons2007a, Drera2011}. Due to the low oxygen partial pressure during growth oxygen vacancies are expected to form in sample OP. Oxygen vacancies induce effective electron doping and create lattice defects. Feature G$_{1}$ is naturally explained by the filling of the bottom of the \STO\ conduction band. The fact that it is visible on-resonant but not off-resonant shows that the bottom of the conduction band is predominantly of Ti 3d character. The relative intensity of G$_{1}$ and G$_{2}$ depends on the specific photon energy, G$_{1}$ resonating more strongly for e$_{g}$ excitations. For \STO\ this has been associated previously with the k$_{z}$ dispersion of the crystalline material. In particular, maxima of G$_{1}$ have been explained by the E$_{F}$ crossing of an electron pocket\cite{Ishida2008}. 
The shape of the low energy spectral weight and the dependence of G$_{1}$ on photon energy is very similar to n-doped \STO\ \cite{Ishida2008}. 
The fact that strong resonance enhancement occurs already before the onset of the main Ti$^{4+}$ absorption (spectrum "pre" in Fig. 2) is consistent with the in-gap states being of Ti$^{3+}$ character because the Ti$^{3+}$ absorption starts at lower energy than Ti$^{4+}$ (see inset Fig. 2c).

Finally, between 13 and 15 eV weak peaks appear on resonant (S$_{1}$ and S$_{2}$) depending on the intermediate state being of t$_{2g}$  or e$_{g}$  symmetry. S$_{2}$ is actually difficult to recognize in Fig. 2a, but becomes apparent when the difference spectra are considered (see Fig. 4). Ti - related intensity in this energy region has been assigned previously to a 3d$^{1}$L$^{2}$ satellite state, where L refers to a ligand hole. Similar satellites are observed for the Ti 2p line as well \cite{Higuchi1998}. For clarity we postpone the discussion of the difference of S$_{1}$ and S$_{2}$ to Fig. 4. 

In Fig. 3 we present equivalent results for sample OR which has been exposed to substantially enhanced oxygen concentration during the growth process (p$_{O2}$ = 5  $\cdot$  10$^{-3}$ mbar). This is supposed to reduce the formation of oxygen vacancies and hence of effective electron doping. The shape of the valence band is similar to sample OP and we apply  the same notation scheme. In agreement with the high resistance of the sample, substantial charging related shifts have been observed. Therefore, the valence bands in Fig. 3 have been aligned to sample OP using the shallow core levels. \cite{ [{This procedure neglects doping dependent energy shifts. However, we are interested here in the direct comparison of features of similar origin.}] Note1} In contrast to OP there exists a low binding energy foot near E$_{F}$ even for the off resonant case. When sweeping the photon energy through the Ti L$_{2,3}$ edge some enhancement occurs in that region, but no intensity is revealed at E$_{F}$ (Fig. 3b). The Ti 3d related enhancement is centered at E$_{B}$ = 1 eV which is at the same position as feature G$_{2}$ for sample OP. Hence, in contrast to OP, G$_{1}$ is absent and the density of states at E$_{F}$ sharply reduced. This is consistent with a reduced n - doping of the sample due to the higher oxygen partial pressure. It is also consistent with the measured resistance.  

Fig. 3c presents a comparison of the Ti L$_{2,3}$ absorption edges of samples OP and OR. The overall shape is  similar, but OP has slightly broader peaks. The latter effect is more clearly seen for the Ti 2p photoemission measurements (Fig. 5) and naturally explained by increased disorder due to the larger density of oxygen vacancies. Also shown in Fig. 3c is the integrated intensity of the main valence band (E$_{B}$ = 3 - 9 eV) as a function of photon energy (red asterisks). This follows the absorption spectra rather well, confirming the Ti$^{4+}$ character of these states.

Apart from the Ti - related features G$_{1}$ and G$_{2}$ a foot structure appears around E$_{B}$  = 2 eV for sample OR. Since this is observed off-resonant it must be O 2p related. The additional oxygen during growth of OR avoids formation of oxygen vacancies but may introduce new types of disorder. It may also cause a less coherent film growth.  

\begin{figure}[h!]
\includegraphics[width=1\linewidth]{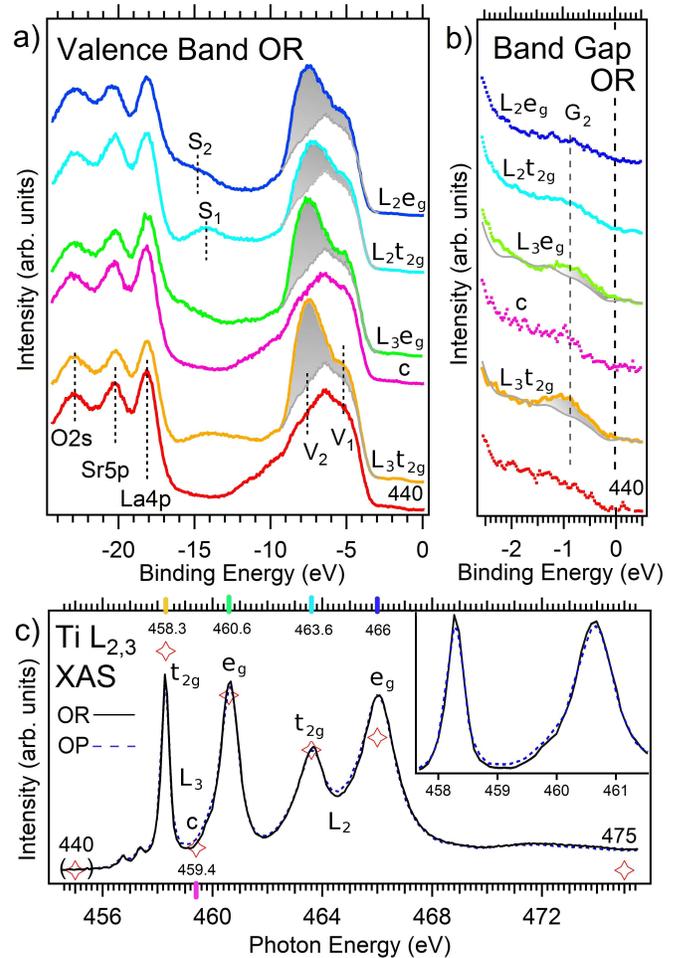}% Here is how to import EPS art
\caption{(Color online) Valence band photoemission of \LAO\ - \STO\ (sample OR, grown in oxygen rich atmosphere). a) Ti 2p $\rightarrow$ 3d resonant photoemission valence band obtained by varying the photon energy through the Ti L$_{2,3}$ absorption threshold. The grey area shows the enhancement of Ti 3d derived states by comparison with off-resonant spectra at 440 eV. b) Expanded view on the spectra shown in (a) near E$_{F}$. Grey line corresponds to the off-resonant measurement. c) Ti L$_{2,3}$ absorption edges of samples OR and OP taken in total electron yield mode. Color bars indicate the photon energy for the valence band spectra. Red asterisks correspond to the integrated valence band between E$_{B}$ = 3 and 9 eV for different photon energies (constant initial state scan - CIS). Inset: Expanded view on the first two peaks.} \label{fig:Fig2}
\end{figure}

Fig. 4 provides a direct comparison of the on-resonant valence bands of both samples. Spectra have been normalized at the O 2s peak and background corrected. Difference spectra (PDOS) have been rescaled to correct for the different numbers of \LAO\ layers \cite { [{The rescaling factor was calculated according to I$_{OP}$/I$_{OR}$ = exp [-d$_{u.c.}$/$\lambda$], where d$_{u.c.}$ = 3.81 \AA\ is the 	dimension of one \LAO\ unit cell and $\lambda$ =11.3 \AA\ is the electron escape length.}] Note}. We start the discussion with the main valence band (features V$_{1}$ and V$_{2}$). The overall shape of the main valence band agrees well with the calculated Ti 3d density of states (DOS) of \STO\ shown at the bottom of Fig. 4a. Interestingly, the relative intensity of V$_{2}$ is reduced for sample OP compared to OR. OP is the sample with more oxygen vacancies and V$_{2}$ originates from bonding Ti 3d - O 2p states. This suggests, that the vacancies disturb the crystalline order and chemical bonding.
Also the reduced intensity of sample OR directly at E$_{F}$ becomes apparent. The intensity around E$_{B}$ = 1 eV (corresponding to feature G$_{2}$ in Fig. 2 + 3) is larger for sample OR when normalized to the shallow core levels (Fig. 4b, upper spectra) due to the additional foot structure which is absent for sample OP. The Ti 3d related intensity at E$_{B}$ = 1 eV, however, is comparable for both samples indicating a similar concentration of Ti defects. 

In Fig. 4a the different position of S$_{1}$ and S$_{2}$ for t$_{2g}$ and e$_{g}$ excitations is clearly visible. We have assigned these satellites to a 3d$^{1}$L$^{2}$ configuration. However, the d - electron of this excited state may occupy t$_{2g}$ or e$_{g}$ levels, which differ in energy by the crystal field splitting. Indeed the energy difference between S$_{1}$ and S$_{2}$ agrees with the t$_{2g}$ - e$_{g}$ distance obtained from the XAS peak distance. Moreover, the e$_{g}$ - satellite should appear at higher binding energy than the t$_{2g}$ - satellite, which is indeed the case.

\begin{figure}[h!]
\includegraphics[width=1\linewidth]{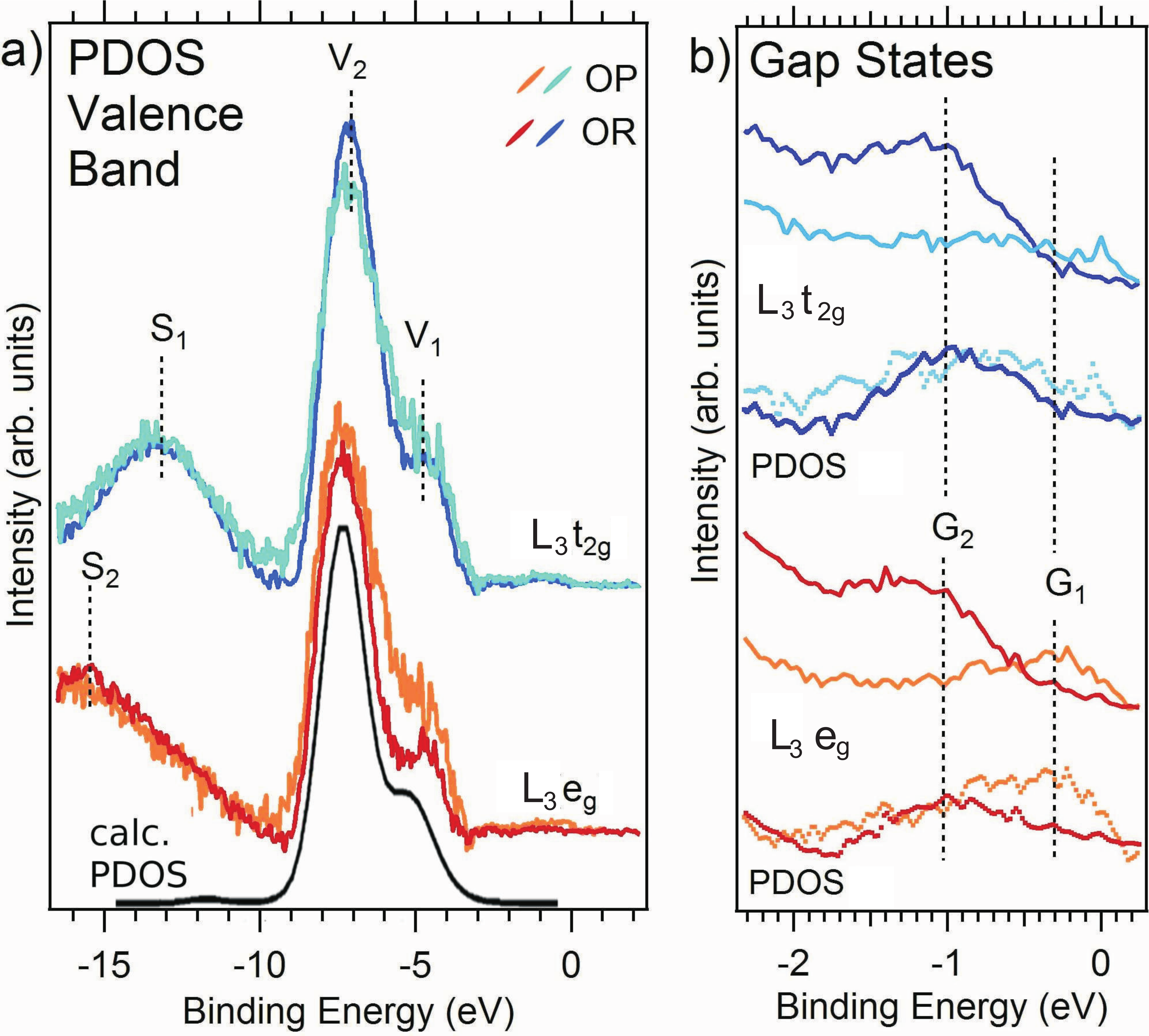}% Here is how to import EPS art
\caption{(Color online) a) Resonance enhancement of the valence band for h$\nu$ = 458.3 eV (t$_{2g}$) and 460.6 eV (e$_{g}$) for samples OP and OR. The calculated density of Ti 3d states is reproduced from Fujii et al. \cite{Fujii2005}. Shown is the difference between on- and off-resonant. b) upper part (blue spectra): Comparison of low energy spectra (uppermost curves) and resonance enhancement (difference on-, off- resonant) for h$\nu$ = 458.3 eV excitation energy (t$_{2g}$). Lower part (orange/red): The same for h$\nu$ = 460.6 eV (e$_{g}$).}\label{fig:Fig3}
\end{figure}

The description of the valence band data is consistent with the assumption, that the concentration of oxygen vacancies is considerably different in the samples and that these defects cause an electron doping of the sample.

\subsection{Core Levels}

The presence of  charge carriers of Ti 3d$^{1}$ character must also alter the shape of the Ti core levels for they effectively correspond to Ti$^{3+}$ states, which differ in energy from the host Ti$^{4+}$ environment. In particular Ti$^{3+}$ shows up at the low energy side of the Ti 2p core level as has been quantitatively evaluated by Sing et al. \cite{Sing2009}. 

In Fig. 5 we present Ti 2p spectra for both samples measured around 1200 eV and 1500 eV photon energy. A Shirley background has been subtracted from the raw data. For the OP 1210 eV spectrum the contribution of an Auger line has been subtracted out before the background subtraction\cite{[{It is the La M$_{4}$N$_{4,5}$N$_{6,7}$ transition at E$_{kin}$ = 740 eV that partly overlaps the Ti 2p emission.  }] Oh1988}. The spectra have been aligned to the Ti 2p$_{3/2}$ line. No reference is made to absolute binding energies due to obvious charging related shifts of sample OR and occasional residual shifts of sample OP. 

For sample OP a low energy foot is discernible at the Ti 2p$_{3/2}$ line, which is absent or much weaker for sample OR. Each spectrum has been fitted with two doublets of Voigt functions. The intensity ratio of the generic Ti$^{4+}$ peaks has been fixed to 2 : 1, whereas the second, smaller component was allowed to vary freely.  For all spectra the presence of the second component improves the fit significantly. The second component corresponds to Ti$^{3+}$ states, which has been also observed in the valence band near  E$_{F}$ on resonance (see above). 
The intensity ratio of the Ti$^{3+}$ doublet is close to 2 : 1 for OR but not for OP. For the 1210 eV OP spectra the Ti$^{3+}$ 2p$_{1/2}$ peak shows an artificially broad and intense structure, which might be related to incomplete background removal. The Ti$^{3+}$ 2p$_{1/2}$ component for OP 1500 eV, on the other hand, is rather sharp and has low intensity, possibly related to the signal to noise ratio at hand. Nevertheless, clear observations can be made from Fig. 5. For sample OP the energy position of the Ti$^{3+}$ doublet is considerably shifted to lower energies with respect to OR. The absolute values of the Ti$^{3+}$ 2p$_{3/2}$ - Ti$^{4+}$ 2p$_{3/2}$ energy separation for OP (ca. 2.0 eV) are close to previous reports for similar samples \cite{Sing2009}.

\begin{figure}[h!]
\includegraphics[width=1\linewidth]{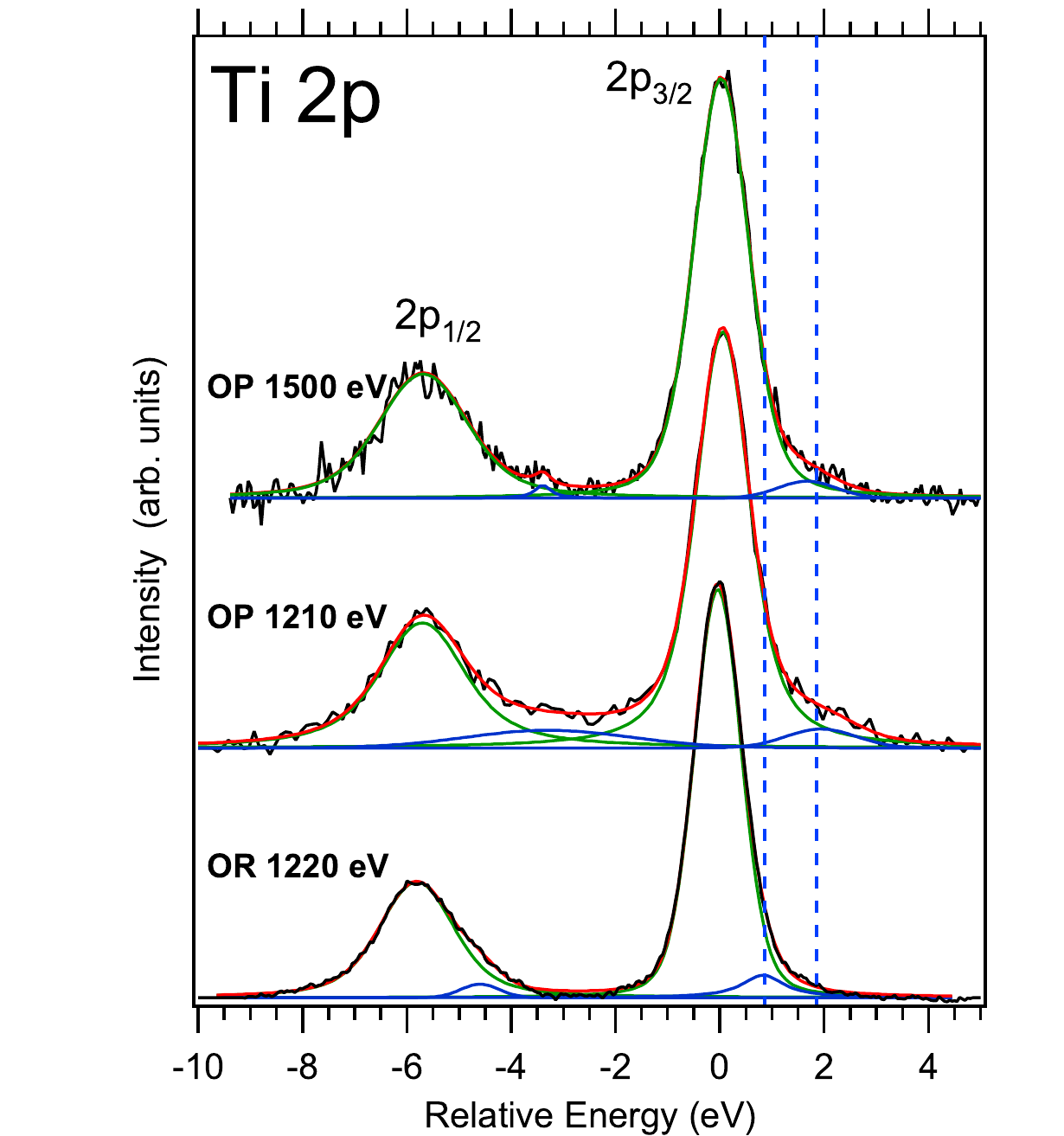}% Here is how to import EPS art
\caption{(Color online) Ti 2p spectra measured in normal emission geometry fitted with Voigt functions (see text). Ti$^{4+}$ peaks are shown in green, the smaller Ti$^{3+}$ peaks in blue. The overall fit is red. Background has been removed before fitting. Vertical dashed lines highlight the positions of the low energy components.}\label{fig:Fig4}
\end{figure}

The appearance of two distinct Ti$^{3+}$ states in the Ti 2p spectra is fully consistent with the observations from the valence band where also two Ti$^{3+}$ components have been found. We associate the Ti$^{3+}$ components of sample OP with structure G$_{1}$ (\STO\ conduction band states) and Ti$^{3+}$ of sample OR with G$_{2}$. Since G$_{2}$ is present for both samples it contributes to the OP sample as well but is masked by G$_{1}$. Only when G$_{1}$ is absent (sample OR) it can be resolved by fitting. The results of the fitting are summarized in table I.  We use the  Ti$^{3+}$/Ti$^{4+}$  value for the following estimation of the charge carrier density near the interface.

 \begin{table}
 \caption{\label{tab:Table1} Fit parameter for Fig. 5
 .}
 \begin{ruledtabular}
 \begin{tabular}{cccc}
Sample & FWHM 2p$_{3/2}$ & FWHM 2p$_{1/2}$ & I(Ti$^{3+}$)/I(Ti$^{4+}$)\\
\hline
OR 1220 eV & 1.07 & 1.70 & 0.068 $\pm$ 0.019 \\
OP 1210 eV & 1.22 & 1.94 & 0.047 $\pm$ 0.04\\
OP 1500 eV & 1.24 & 2.0 & 0.038 $\pm$ 0.017\\
 \end{tabular}
 \end{ruledtabular}
 \end{table}

The electron escape length $\lambda$ of the Ti 2p photoelectrons for h$\nu$ = 1200 eV in \LAO\ is calculated by using the TTP - 2M formalism to be 15.6 \AA\ or approximately 4 unit cells\cite{Tanuma1991}. By definition the majority of the photoemission signal originates from a layer of this thickness. Note, that for the Ti 2p photoelectrons the \LAO\ overlayer acts as damping barrier, it does not change the Ti$^{3+}$/Ti$^{4+}$ ratio - provided Ti intermixing can be neglected. Assuming a constant Ti$^{3+}$ distribution over $\lambda$, the charge carrier concentration for sample OP near the interface is appr. 4 - 5 \%. The exact value depends on the total thickness of the Ti$^{3+}$ layer d, which, unfortunately, cannot be obtained from the present experiment with sufficient accuracy. A lower limit of the sheet carrier density based on the above assumptions yields n $\approx$ 1.3 $\cdot$ 10$^{14}$ cm$^{-2}$. Note, that the true value could be substantially larger if d $\gg$ $\lambda$. 

The line widths of the Ti$^{4+}$ peaks of sample OR are narrower than for sample OP (see Table 1). This is consistent with the Ti L absorption lines. It is probably related to increased overall disorder at the Ti - site in sample OP due to the higher density of oxygen vacancies with respect to OR. 

\begin{figure}[h!]
\includegraphics[width=1\linewidth]{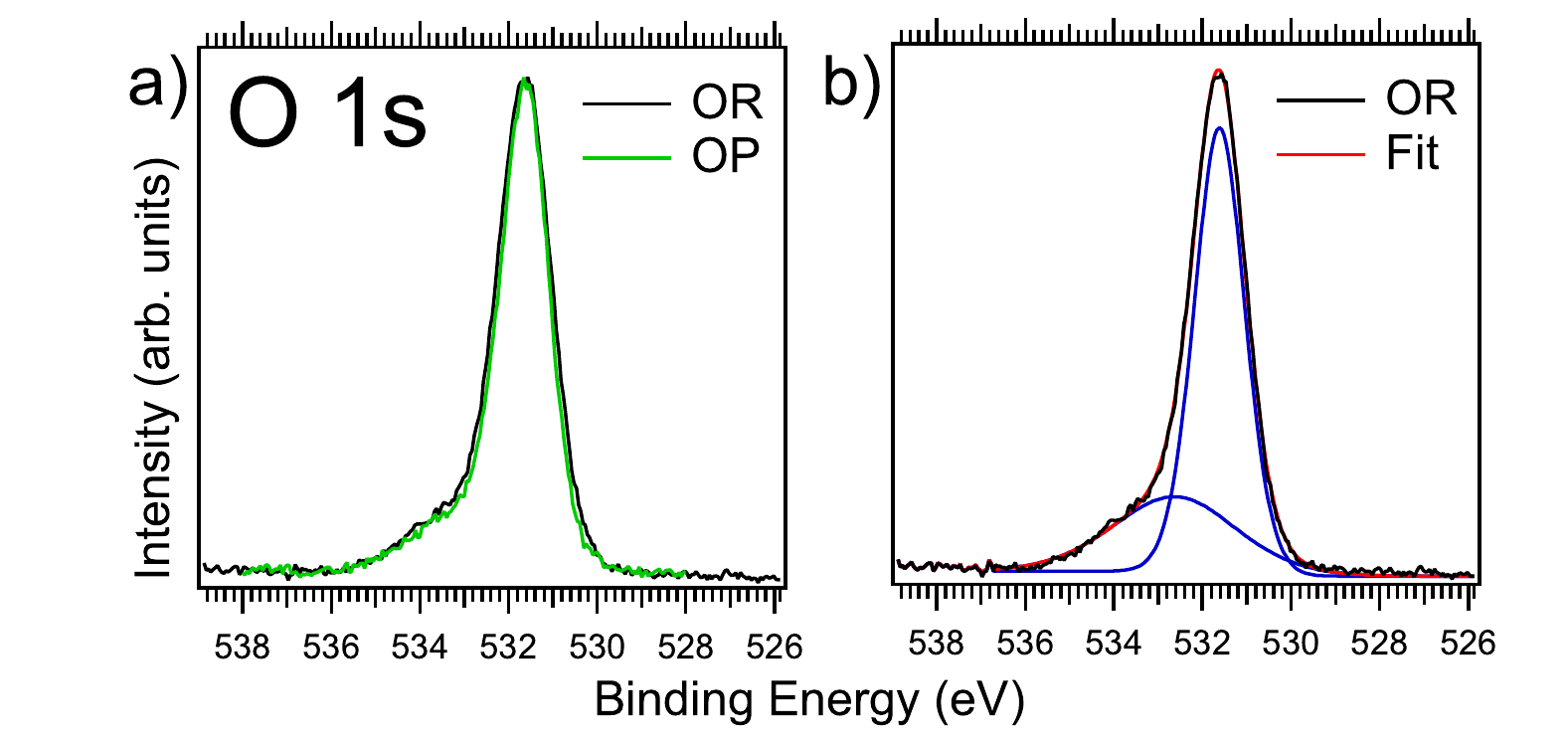}% Here is how to import EPS art
\caption{(Color online) O 1s spectra measured in normal emission geometry with h$\nu$ = 1200 eV . }\label{fig:Fig5}
\end{figure}

Fig. 6 shows the O 1s lines for both samples. Whereas the Ti 2p measurement is mainly sensitive to the interface region, O 1s probes predominantly the \LAO\ surface. The line consists of two components: the main component at lower binding energy stemming from the \LAO\ lattice, and the contamination shoulder at higher binding energies, probably related to hydroxylation \cite{Chambers2001}. 
It has been pointed out recently that adsorbates at the polar \LAO\ surface may alter the interface properties by modifying the electrostatic potential across the \LAO\ overlayer \cite{Xie2011}. In this respect it is important to check whether the surfaces of our two samples look substantially different concerning the adsorbates. As can be seen in Fig. 6a this is not the case. The spectra are almost identical. This rules out the possibility that the observed differences among the samples arise from different surface characteristics. The intensity of the contamination peak is $\sim$ 30 \%, similar to previous reports \cite{Qiao2011}.

\section{Discussion}

The oxygen treatment has apparent consequences for the electronic structure of the \LAO\  - \STO\ interface. We focus the discussion on the in-gap features as they determine the transport properties. We have identified three types of in-gap states: G$_{1}$, G$_{2}$ and the foot structure at the valence band onset of sample OR. G$_{1}$ and G$_{2}$ are related to Ti 3d$^{1}$ as they resonate across the Ti L edge. They are ubiquitous for \STO\ and related systems and have been observed for thin films \cite{Ishida2008} and, similarly, single crystals \cite{Aiura2002}. Their origin is at the center of a longstanding and, so far, not completely resolved debate.
The most widespread interpretation is in terms of oxygen vacancy related defects. The feature near E$_{B}$  = 1 eV is then due to Ti 3d - O-vacancy complexes where an additional electron is trapped and forms a localized state. In contrast, the states at E$_{F}$ are formed by electrons injected into the \STO\ conduction band. But if only oxygen vacancies were decisive for the formation of feature G$_{2}$ one would expect some dependence of its intensity on oxygen pressure during growth. From the direct comparison in Fig. 4, however, G$_{2}$ is of similar intensity among the samples, or even more pronounced for sample OR. Therefore, a scenario where G$_{2}$ depends on the oxygen vacancy density alone seems inappropriate. Rather one has to consider defects arising from other types of local chemical disorder. 

Another difference between the samples is the appearance of the foot structure of sample OR. It is apparently not related to Ti 3d orbitals but rather to oxygen orbitals. If one recalls, that sample OR has been grown in relatively high oxygen pressure compared to sample OP, it is tempting to associate the additional intensity with some additional or disordered oxygen in the lattice. However, the perovskite lattice is not susceptible to, e.g. oxygen interstitials. Rather this feature could be related to a less coherent film growth under higher oxygen pressure.

Coming back to the origin of  G$_{2}$, an often discussed possibility in this regard concerns the cations, La and Sr. It was observed early on that Ar - sputtering of \STO\ surfaces does not only create oxygen vacancies but also strontium vacancies \cite{Henrich1978}. The PLD process relies on a non-equilibrium ion bombardment of the substrate, which, moreover, has been treated beforehand  by a aggressive chemical procedure. Therefore, Sr-vacancies could play a role for \STO. Furthermore, the deposition of the \LAO\ layers is known to introduce partial intermixing across the interface, as has been observed previously by several groups \cite{Qiao2010, Willmott2007, Kalabukhov2007}. This means that Sr diffuses into the \LAO\ layer and La into \STO. However, it has been also reported that the intermixing depends on the oxygen pressure as well \cite {Kalabukhov2011}. This can be understood by deceleration and thermalization of the incoming ions by the more dense oxygen atmosphere. 
This would imply a dependence of  G$_{2}$ on the oxygen pressure during growth. Nevertheless, the occurrence of such defects is related to the PLD process and cannot be ruled out.

We mention that alternative interpretations of G$_{2}$ have been proposed, i.e. such that do not involve defects. In the limit of high doping levels the system approaches a 3d$^{1}$ configuration, which results in a Mott insulating state, as is realized e.g. in LaTiO$_{3}$ \cite{Fujimori1992}. It is clear that the spectral function has to undergo drastic changes in this process. In this respect G$_{2}$ could be associated with a precursor of the lower Hubbard band. In fact the in-gap states of LaTiO$_{3}$ -  \STO\ interfaces have been studied by Takizawa et al. by off- and on-resonant photoemission and found to be consistent with such an interpretation \cite{Takizawa2006}. However, the doping levels for our samples, especially OR, are too small to sustain this scenario.

Ishida et al. \cite{Ishida2008} assigned the in-gap state at E$_{B}$ = 1.5 eV in Nb-doped \STO\ to a local screening channel of the d - orbitals. However, in our case the potential well screened state at E$_{F}$ is virtually absent for sample OR, rendering this scenario for our situation unlikely as well. 

From our point of view a defect related origin of the G$_{2}$ peak remains as the most plausible possibility. But it is not easy to clarify its exact nature. This issue clearly requires further investigations. 

We now turn to another open question.
Within the picture of the polar catastrophe charge carriers should be present at the interface of sample OR, but are, within the sensitivity of our experiments, not observed. The resistance is too high and no intensity is measured at E$_{F}$. 
Why is this so? We propose a scenario where the G$_{2}$  defects (which are of substantial Ti 3d$^{1}$ character) cause a localization of electrons and deprive thereby the interface of itinerant charge carriers. The additional charge transfered to the interface by the polar catastrophe reconstruction maybe lost in this way for electrical conductivity, although it still balances the electrostatics at the interface and may give rise to localized magnetic moments \cite{Brinkman2007}. For sample OP there are enough oxygen vacancy related additional electrons to compensate this and still populate the conduction band, but not for sample OR.
In fact the defect density is high enough to support this scenario: i)  From the direct comparison in Fig. 4 and from the fitting results of the Ti 2p lines the conclusion is justified, that G$_{1}$ and G$_{2}$ have roughly the same intensity, i.e. considering the reasonable transport properties of OP, the defect density must be substantial. ii) From the Ti$^{3+}$ component of the OR - Ti 2p line a minimal sheet defect density of $\sim$ 2 $\cdot$ 10$^{-14}$ cm$^{-2}$ near the interface can be estimated (table I). This could be enough to outweigh the effect of the polar catastrophe with an upper limit of the sheet carrier density of 3.4 $\cdot$ 10$^{-14}$ cm$^{-2}$.

Alternatively, the polar discontinuity maybe mitigated by cationic displacement and intermixing from the beginning, leaving the presence of oxygen vacancies as the only source of charge carriers . 

\section{Summary}

We have investigated the \LAO\ - \STO\ heterointerface by soft x-ray spectroscopy for different sample treatments. The population of the \STO\ conduction band is directly seen for the sample grown in low oxygen atmosphere but is absent for the sample grown in high oxygen atmosphere. Another prominent Ti - related state is observed within the \STO\ gap and is probably related to disorder induced defects. Both types of Ti$^{3+}$ related states - conduction electrons and defects - are visible near the Fermi energy but also at the low energy side of the Ti 2p core level. The results signal a varying charge carrier concentration among the samples, likely caused by a growth dependent defect density.

\bibliography{LAOSTO}
\bibliographystyle{apsrev4-1}

\end{document}